\documentclass[apj]{emulateapj}
\usepackage{graphicx}

\newcommand{\ir}{\mbox{\rm IR}} 
\newcommand{\co}{\mbox{\rm CO}} 
\newcommand{\hi}{\mbox{\rm \ion{H}{1}}} 
\newcommand{\hii}{\mbox{\rm \ion{H}{2}}}
\newcommand{\htwo}{\mbox{\rm H$_2$}} 
\newcommand{\ha}{\mbox{\rm H$\alpha$}} 
\newcommand{\kmpers}{\mbox{km~s$^{-1}$}}
\newcommand{\xcounits}{\mbox{cm$^{-2}$ (K km s$^{-1}$)$^{-1}$}}
\newcommand{\xco}{\mbox{$X_{\rm CO}$}}
\newcommand{\htwosd}{\mbox{$\Sigma_{\rm H2}$}}
\newcommand{\Msunperpc}{\mbox{\rm M$_{\odot}$ pc$^{-2}$}}

\newcommand{\SFRsd}{\mbox{$\Sigma_{\rm SFR}$}}

\newcommand{\ergpers}{\mbox{erg~s$^{-1}$}}

\shorttitle{Scale Dependence of $\tau_{\rm dep}$ in M33}
\shortauthors{Schruba et al.}
\slugcomment{Accepted for publication in \emph{The Astrophysical Journal}}

\begin{document}
\title{The Scale Dependence of the Molecular Gas Depletion Time in M33}

\author{Andreas Schruba\altaffilmark{1}, Adam Leroy\altaffilmark{2,1,4}, Fabian Walter\altaffilmark{1}, Karin Sandstrom\altaffilmark{1}, Erik Rosolowsky\altaffilmark{3}}
\altaffiltext{1}{Max-Planck-Institut f\"ur Astronomie, K\"onigstuhl 17, D-69117 Heidelberg, Germany}
\altaffiltext{2}{National Radio Astronomy Observatory, 520 Edgemont Road, Charlottesville, VA 22903, USA}
\altaffiltext{3}{University of British Columbia Okanagan, 3333 University Way,
Kelowna, BC V1V 1V7, Canada}
\altaffiltext{4}{Hubble Fellow}

\begin{abstract}
We study the Local Group spiral galaxy M33 to investigate how the
observed scaling between the (kpc-averaged) surface density of
molecular gas ($\Sigma_{\rm H2}$) and recent star formation rate
($\Sigma_{\rm SFR}$) relates to individual star-forming regions. To do
this, we measure the ratio of CO emission to extinction-corrected
\ha\ emission in apertures of varying sizes centered both on peaks of CO
and H$\alpha$ emission. We parameterize this ratio as a molecular gas
(\htwo) depletion time ($\tau_{\rm dep}$). On large (kpc) scales, our
results are consistent with a molecular star formation law ($\SFRsd
\sim \Sigma_{\rm H2}^b$) with $b\sim 1.1 - 1.5$ and a median
$\tau_{\rm dep} \sim 1$~Gyr, with no dependence on type of region
targeted. Below these scales, $\tau_{\rm dep}$ is a strong function of
adopted angular scale and the type of region that is targeted. Small
($\lesssim 300$~pc) apertures centered on CO peaks have very long
$\tau_{\rm dep}$ (i.e., high CO-to-H$\alpha$ flux ratio) and small
apertures targeted toward H$\alpha$ peaks have very short $\tau_{\rm
  dep}$. This implies that the star formation law observed on kpc scales
breaks down once one reaches aperture sizes of $\lesssim 300$~pc.
For our smallest apertures ($75$~pc), the difference in $\tau_{\rm dep}$
between the two types of regions is more than one order of magnitude.
This scale behavior emerges from averaging
over star-forming regions with a wide range of CO-to-H$\alpha$ ratios
with the natural consquence that the breakdown in the star formation law
is a function of the surface density of the regions studied. We
consider the evolution of individual regions the most likely driver
for region-to-region differences in $\tau_{\rm dep}$ (and thus the
CO-to-H$\alpha$ ratio).
\end{abstract}


\section{Introduction}
\label{sec:intro}

The observed correlation between gas and star formation rate surface
densities (the `star formation law') is one of the most widely used
scaling relations in extragalactic astronomy
\citep[e.g.,][]{Schmidt1959, Kennicutt1998}.  However, its connection
to the fundamental units of star formation, molecular clouds and young
stellar clusters, remains poorly understood. On the one hand, averaged
over substantial areas of a galaxy, the surface density of gas
correlates well with the amount of recently formed stars
\citep[e.g.,][]{Kennicutt1998}.  On the other hand, in the Milky Way
giant molecular clouds (GMCs), the birthplace of most stars, and
\hii\ regions, the ionized ISM regions around (young) massive stars,
are observed to be distinct objects. While they are often found near
one another, the radiation fields, stellar winds, and ultimately
supernovae make \hii\ regions and young clusters hostile to their
parent clouds on small ($\sim$$10$~pc) scales. Thus, while correlated
on galactic scales, young stars and molecular gas are in fact
anti-correlated on very small scales. The details of the transition
between these two regimes remain largely unexplored \citep[though
  see][]{Evans2009}.

Recent observations of nearby galaxies have identified a particularly
tight correlation between the distributions of \emph{molecular} gas
(\htwo) and recent star formation on $\sim$kpc scales
\citep{Murgia2002, Wong2002, Kennicutt2007, Bigiel2008, Leroy2008,
  Wilson2008}.  While the exact details of the relation are still
somewhat uncertain, in the disks of spiral galaxies the
parametrization seems to be a power law, $\Sigma_{\rm SFR} = a
\Sigma_{\rm H2}^b$, with power law index $b \approx 1 - 1.7$ and
coefficient $a$ corresponding to \htwo\ depletion times of $\sim$$2$~Gyr
in normal spirals. 

Both parts of this relation, the surface densities of \htwo\ and
recent star formation, resolve into discrete objects: GMCs and
\hii\ regions, young associations, and clusters. In this paper, we
investigate whether the $\sim$kpc \htwo-SFR relation is a property of
these individual regions or a consequence of averaging over large
portions of a galactic disk (and the accompanying range of
evolutionary states and physical properties). To do so, we compare CO
and extinction-corrected H$\alpha$ at high spatial resolution in the
nearby spiral galaxy M33. We examine how the ratio of CO-to-H$\alpha$
changes as a function of region targeted and spatial scale. M33 is a
natural target for this experiment: it has favorable orientation and
is close enough that peaks in the CO and H$\alpha$ maps approximately
correspond to individual massive GMCs \citep{Rosolowsky2007} and
\hii\ regions \citep{Hodge2002}.

Perhaps not surprisingly, we find that the ratio of \co\ luminosity (a
measure of the molecular gas mass) to extinction-corrected \ha\ flux
(a measure of the star formation rate) depends on the choice of
aperture and spatial scale of the observations. After describing how
we estimate \htwo\ masses and the recent star formation rate
(Section~\ref{sec:data}) and outlining our methodology
(Section~\ref{sec:methodology}), we show the dependence of the
depletion times on spatial scale and region targeted
(Section~\ref{sec:results}). Then we explore physical explanations for
these results (Section~\ref{sec:disc}).

\section{Data}
\label{sec:data}

We require the distributions of \htwo\ and recently formed stars
which we trace via \co\ emission and a combination of
\ha\ and \ir\ emission, respectively.

\subsection{Molecular Gas from \co\ Data}
\label{subsec:moleculargas}

\begin{figure*}
\plottwo{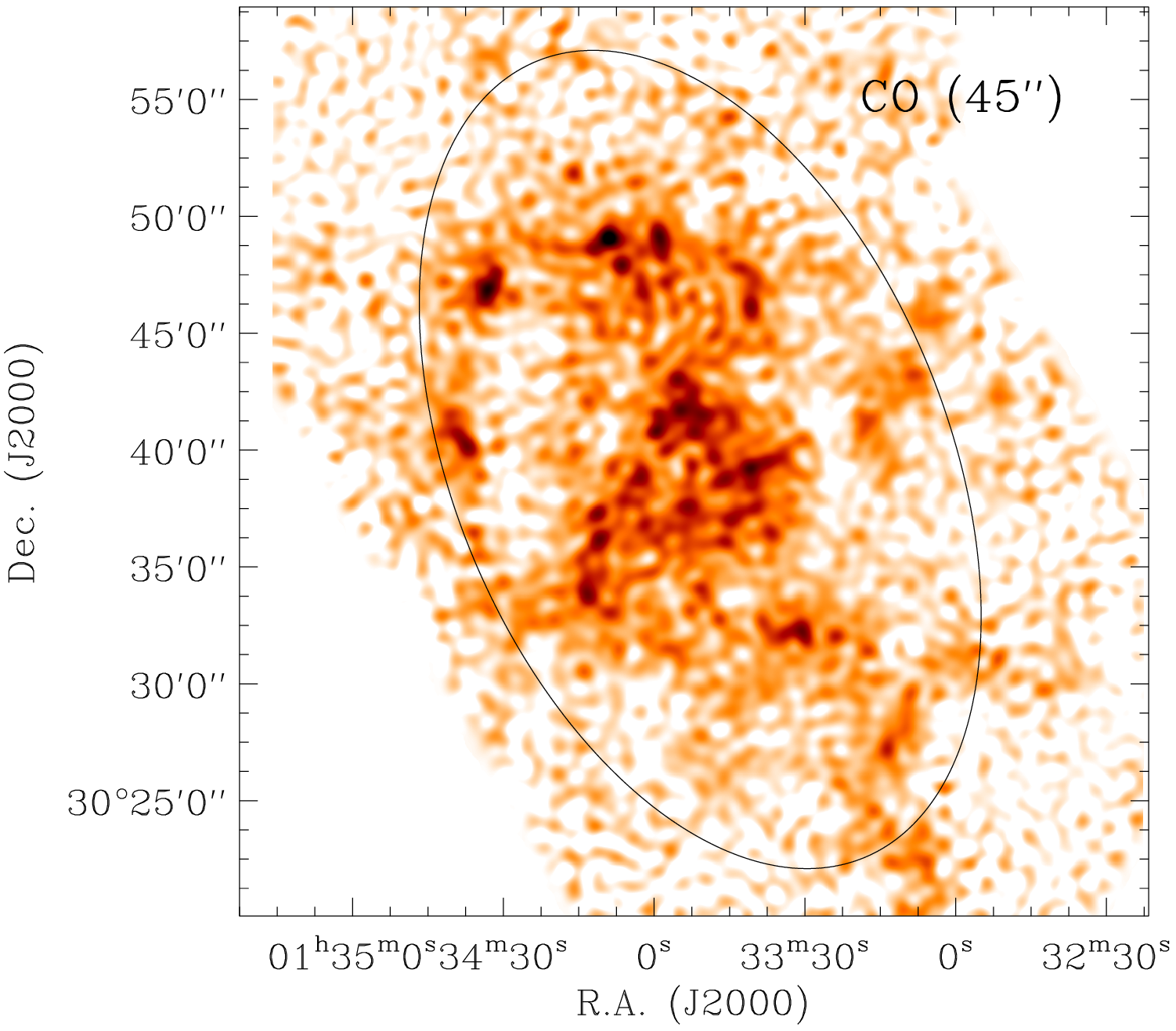}{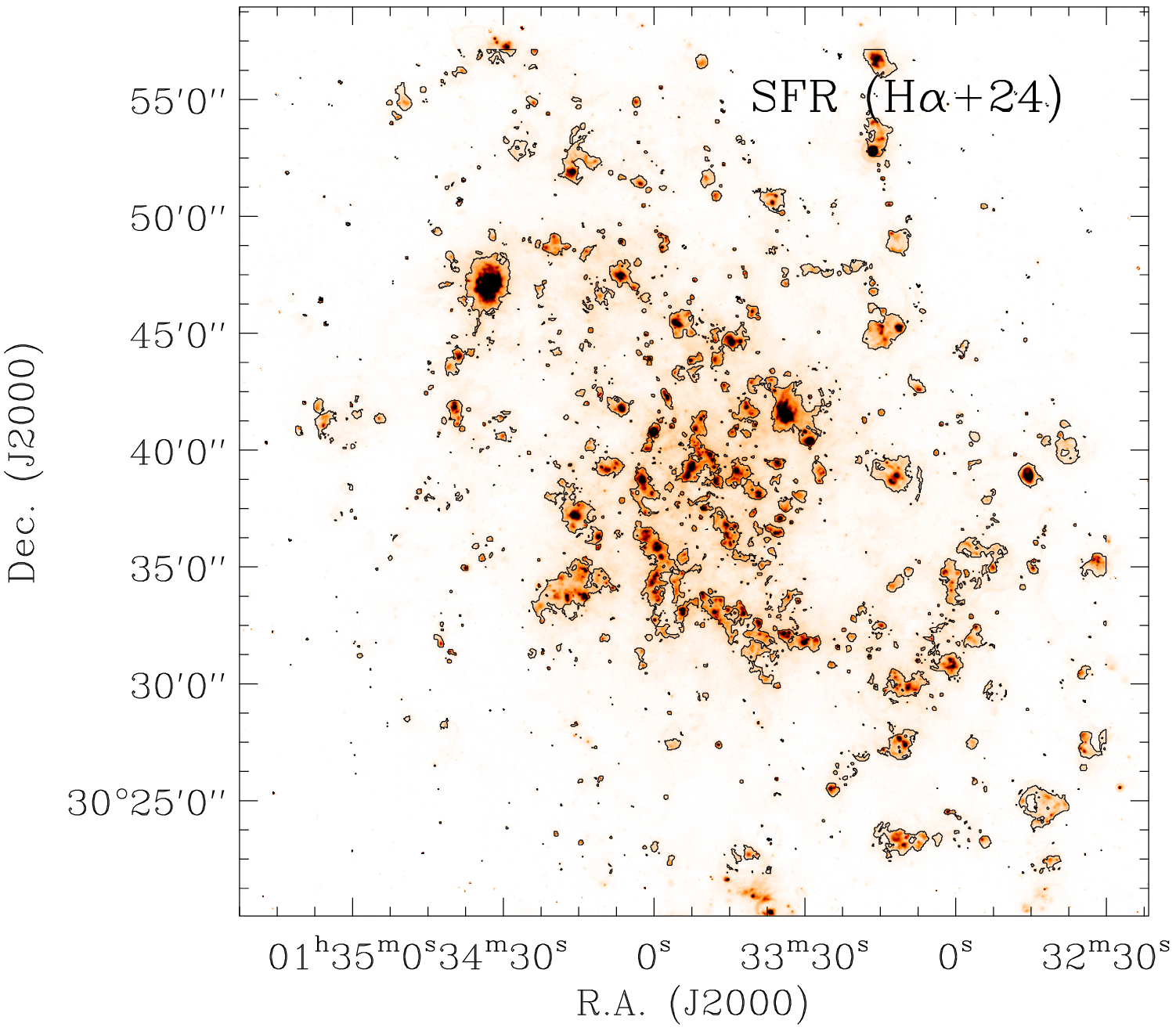}
\plottwo{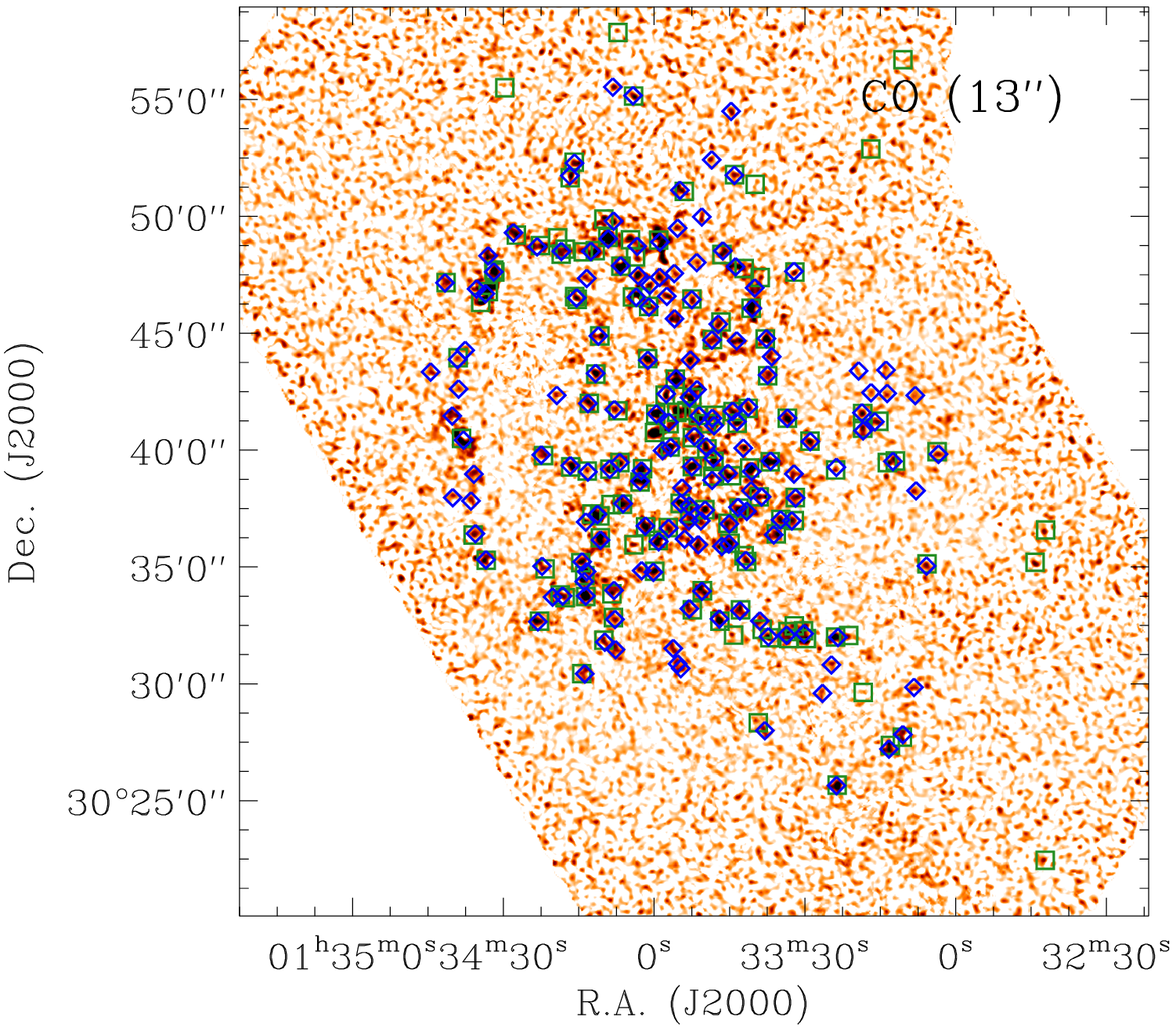}{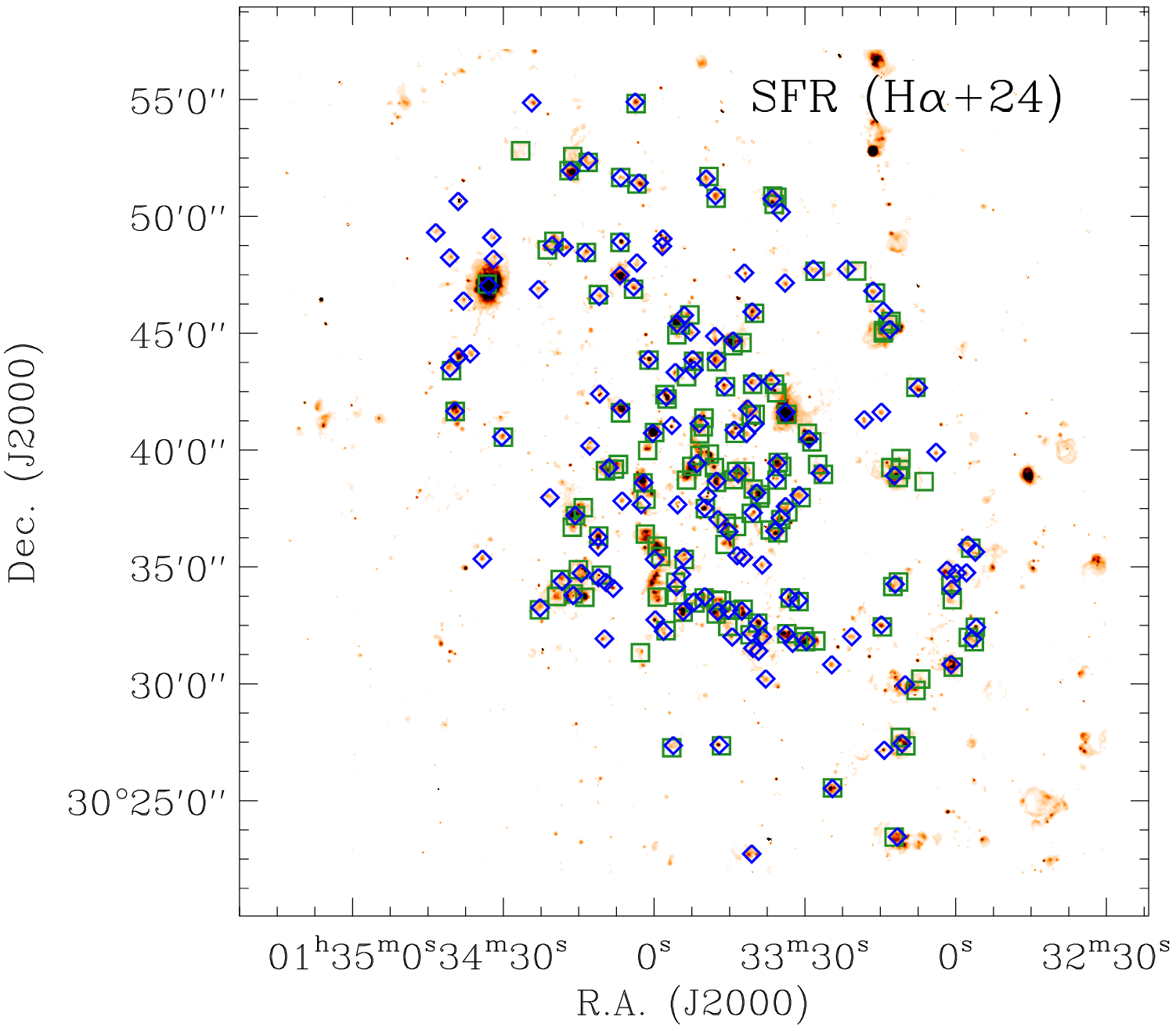}
\plottwo{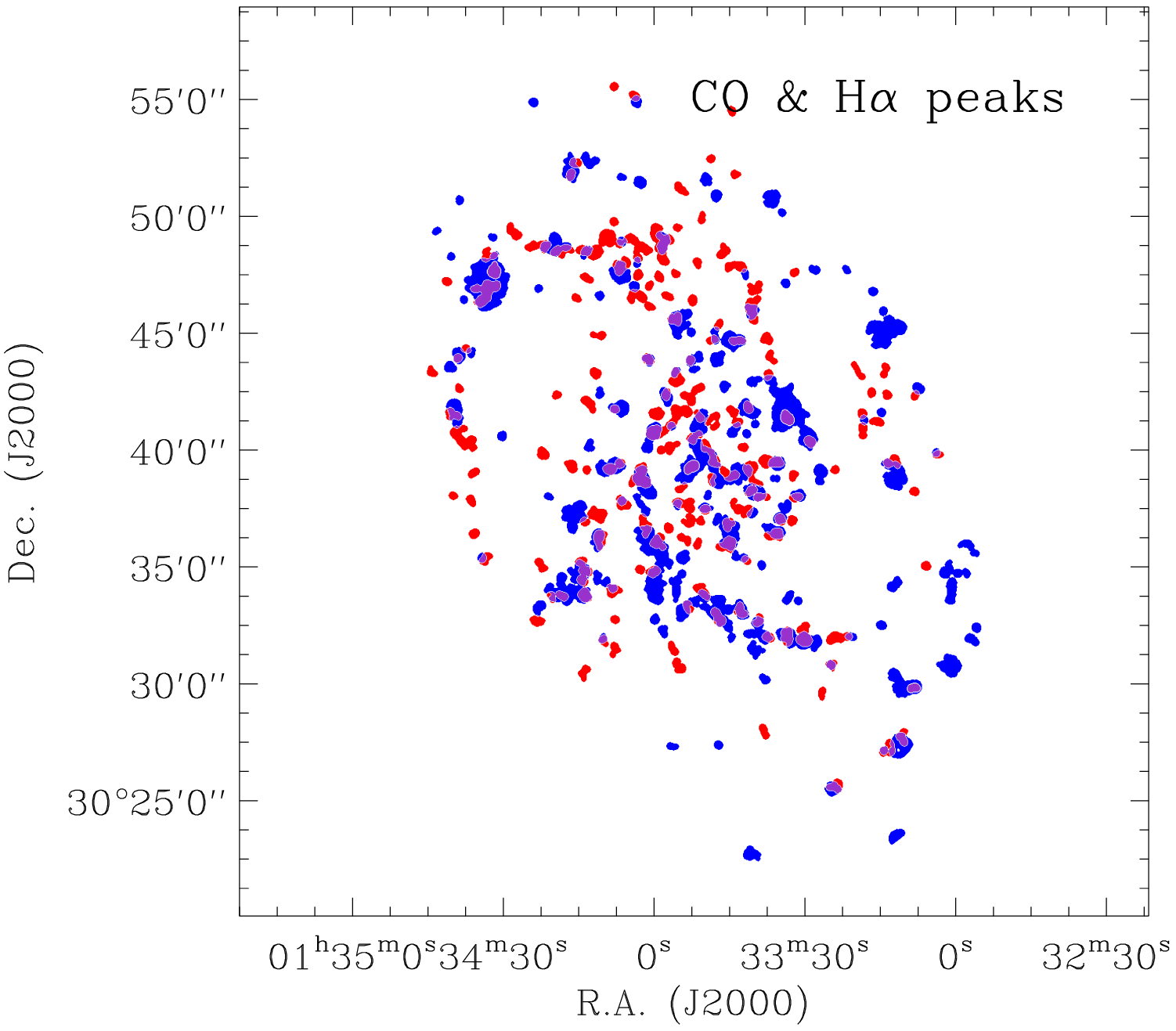}{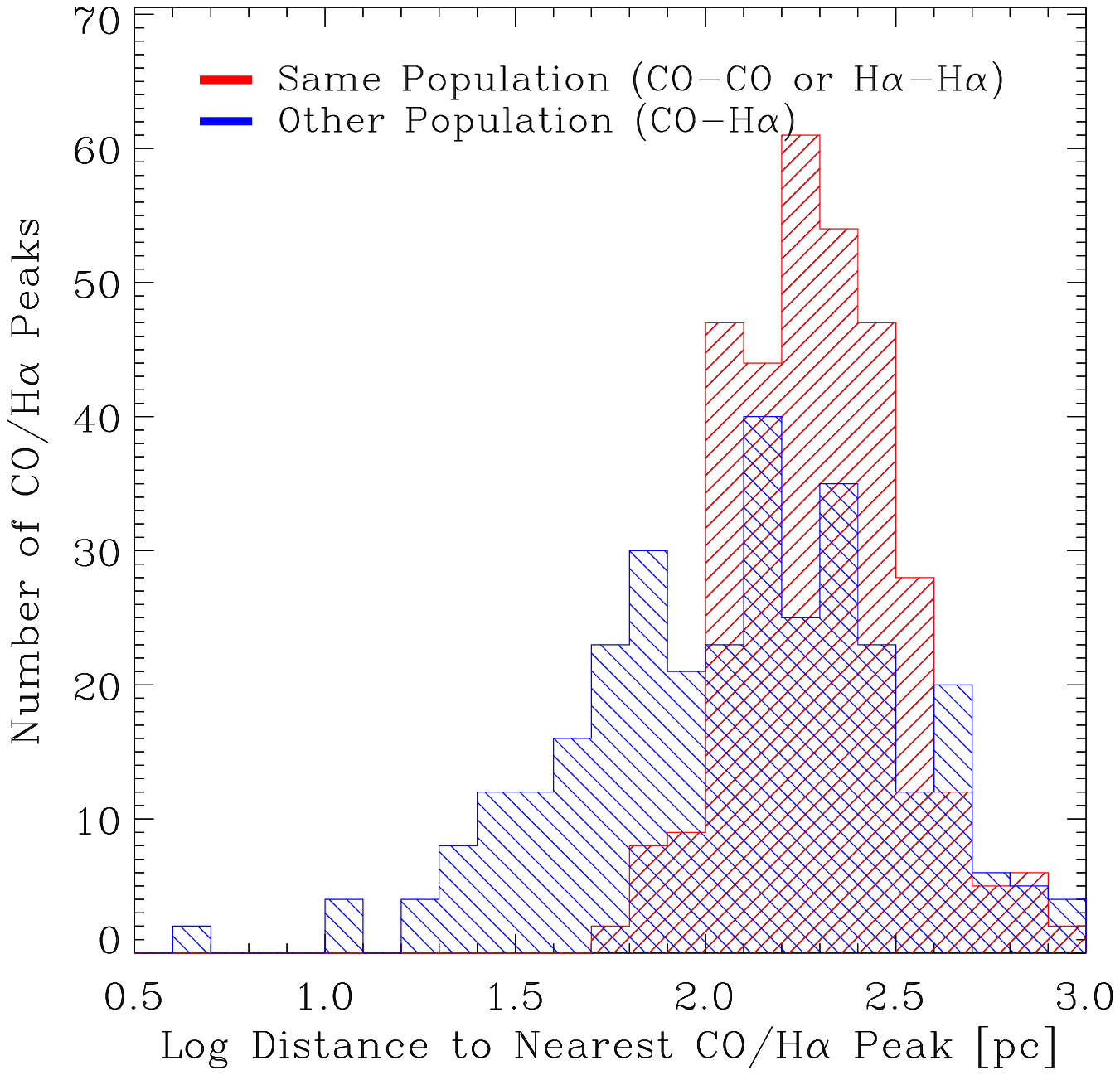}
	\caption{\label{f1} CO and recent star formation in M33. {\em
            Top left:} Masked, integrated CO intensity smoothed to
          $\sim$$45\arcsec$ resolution to enhance the SNR.  The black
          ellipse shows $R_{\rm gal} = 4.5$ kpc ($\sim$$0.6~r_{25}$);
          we carry out our analysis inside this radius. {\em Top
            right:} H$\alpha$ emission corrected for extinction via
          combination with mid-IR emission. A black contour outlines
          regions that remain after the subtraction of the diffuse
          ionized gas (see text). {\em Middle left:} Positions of our
          CO peaks (blue diamonds) plotted on the full resolution CO
          map along with the cataloged positions of GMCs from
          \citet[][green squares]{Rosolowsky2007}. {\em Middle Right:}
          Positions of our H$\alpha$ peaks (blue diamonds) along with
          the $150$ most luminous \hii\ regions cataloged by
          \citet[][green squares]{Hodge2002}. {\em Bottom left:}
          Relative distributions of bright CO (red) and H$\alpha$
          (blue) emission and the area of overlap (purple), inflated
          by $6$\arcsec\ for display reasons.  {\em Bottom right:}
          Histograms of distance from each peak to the nearest peak of
          the same type (red, i.e., H$\alpha$ to H$\alpha$ or CO to
          CO) and of the other type (blue, CO to H$\alpha$ and vice
          versa).}
\end{figure*}

Star-forming clouds consist mainly of \htwo, which cannot be directly
observed under typical conditions. Instead, \htwo\ is usually traced
via emission from the second most common molecule, \co. We follow this
approach, estimating \htwo\ masses from the \co\ $J=1-0$ data of
\citet{Rosolowsky2007}, which combines the BIMA (interferometric) data
of \cite{Engargiola2003} and the FCRAO $14$~m (single-dish) data of
\citet{Heyer2004}. The resolution of the merged data cube is
$13\arcsec \times 2.03$ \kmpers\ with a median $1\sigma$ noise of
$240$~mK ($\sim$$2.1$~\Msunperpc\ for our adopted \xco ).
\citet{Rosolowsky2007} showed that this combined cube recovers
the flux of the \citet{Heyer2004} FCRAO data.

We convert integrated
\co\ intensities into molecular gas surface densities assuming \xco =
$2.0 \times 10^{20}$ \xcounits . This is approximately the Milky Way
conversion factor and agrees well with work on M33 by
\citet[][]{Rosolowsky2003}.  For this \xco :

\begin{equation}
\label{eq:xco}
	\Sigma_{\rm H2}  \left[ {\rm M}_\odot~{\rm pc}^{-2}\right] = 
	4.4~I_{\rm CO} \left[ {\rm K}~{\rm km~s}^{-1}\right]~,
\end{equation}

\noindent where $I_{\rm CO}$ is the integrated CO intensity over the
line of sight and $\Sigma_{\rm H2}$ is the mass surface density of
molecular gas, including helium.

The data cover a wide bandpass, only a small portion of which contains
the CO line. As a result, direct integration of the cube over all
velocities produces an unnecessarily noisy map. Therefore, we ``mask''
the data, identifying the velocity range likely to contain the CO line
along each line of sight. We integrate over all channels with $\pm 25$
\kmpers\ of the local mean \hi\ velocity \citep[using the data
  from][]{Deul1987}. To ensure that this does not miss any significant
emission, we also convolve the original \co\ cube to
$30$\arcsec\ resolution and then identify all regions above
$3\sigma$ in $2$ consecutive channels. Any region within or near
such a region is also included in the mask. We blank all parts of the
data cube that do not meet either criteria and then integrate along
the velocity axis to produce an integrated \co\ intensity map. Figure
\ref{f1} shows this map at full resolution ({\em middle left}) and
smoothed to $\sim$$45\arcsec$ resolution ({\em top left}) to increase the
SNR and highlight extended emission. The noise in the integrated
intensity map varies with position but typical $1\sigma$ values are
$8 - 10$ \Msunperpc ; the dynamic range (peak SNR) is $\sim$$20$.
The $3\sigma$ mass sensitivity in an individual resolution element
is $\sim$$10^5$~M$_\sun$.

\subsection{Recent Star Formation from \ha\ and \ir\ Data}
\label{subsec:starformation}

We trace the distribution of recent star formation using
\ha\ emission, which is driven by ionizing photons produced almost
exclusively in very young (massive) stars. We account for extinction
by combining H$\alpha$ and infrared (24$\mu$m) emission, a powerful
technique demonstrated by \citet{Calzetti2007} and
\citet{Kennicutt2007, Kennicutt2009}. Assuming continuous star
formation over the past $100$~Myr and studying a set of extragalactic
star-forming regions, \citet{Calzetti2007} found the recent star
formation rate (SFR) to be

\begin{equation}
  \label{eq:sfr}
	{\rm SFR} \left[ {\rm M}_\odot~{\rm yr}^{-1} \right]
        = 5.3 \times 10^{-42} \left[ L \left({\rm
            H}\alpha\right) + 0.031\ L \left(24 \mu {\rm
            m}\right) \right]~,
\end{equation}

\noindent where $L$(\ha) and $L(24 \mu {\rm m}) = \nu L_\nu (24
\mu {\rm m})$ are the luminosities of a region in \ha\ emission and at
$24$\micron, measured in \ergpers.

The assumption of continuous star formation is certainly inapplicable
to individual regions, which are better described by instantaneous
bursts \citep[e.g.,][]{Relano2009}. We only report averages of a large
set ($\sim$$150$) of regions, which together constitute a large part of
M33's total \ha, and argue that this justifies the application of
Equation \ref{eq:sfr} (see Section~\ref{subsec:depletiontimes}). In
any case SFR units allow ready comparison to previous work.

\subsubsection{\ha\ Data}
\label{subsec:hadata}

We use the narrow-band \ha\ image obtained by \citet{Greenawalt1998}
with the KPNO $0.6$~m telescope. The reduction, continuum subtraction,
and other details of these data are described by
\citet{Hoopes2000}. Before combination with the IR map, we correct the
\ha\ map for Galactic extinction using a reddening of $\mbox{E(B-V)} =
0.042$ \citep{Schlegel1998} and a ratio of \ha\ narrow band extinction
to reddening of $\mbox{R(\ha)} = 2.33$.

Studies of M33 and other nearby galaxies find typically $\sim$$40$\%
of the \ha\ emission to come from ``diffuse ionized gas''
\citep[DIG,][]{Hoopes2000, Thilker2002, Hoopes2003, Thilker2005}.  The
origin of this emission is still debated; it may be powered by leaked
photons from bright \hii\ regions or it may arise ``in situ'' from
isolated massive stars. We choose to remove this diffuse component
from the H$\alpha$ map and any discussion of \ha\ emission in the
following analysis refers to this DIG-subtracted map (we assess the
impact of this step in the Appendix). We do so using the following
method from \citet{Greenawalt1998}. We begin by median filtering the
\ha\ map with a $900$~pc kernel. We then identify \hii\ regions as
areas in the original map that exceed the median-filtered map by an
emission measure of $50$~pc~cm$^{-6}$ (outlined by black contours in
the top right panel of Figure \ref{f1}). We blank these regions in the
original map and smooth to get an estimate of DIG emission towards the
\hii\ regions. Our working map consists of only emission from the
\hii\ regions after the DIG foreground has been subtracted. The
integrated \ha\ flux (inside $R_{\rm gal} = 4.5$ kpc) allocated to the
diffuse map is $1.05 \times 10^{40}$ \ergpers\ ($44$\%) while the
part allocated to the DIG-subtracted, \hii -region map is
$1.35 \times 10^{40}$ \ergpers\ ($56$\%), in good agreement with
previous results on M33 and other nearby galaxies.

\subsubsection{\ir\ Data}
\label{subsec:irdata}

We measure \ir\ intensities from $24$~\micron\ maps obtained by the
\emph{Spitzer} Space Telescope (PI: \citealt{Gehrz2005}, see also
\citealt{Verley2007}). The data were reduced by K. Gordon (2009,
private communication) following \citet{Gordon2005}.  {\em Spitzer}'s
point spread function at $24$~\micron\ is $\sim$$6$\arcsec, well below
our smallest aperture size ($\sim$$18$\arcsec ) and so is not a large
concern.

As with the \ha\ image, the $24$~\micron\ map includes a substantial
fraction of diffuse emission --- infrared cirrus heated by an older
population, emission from low-mass star-forming regions, and dust
heated leakage from nearby \hii\ regions, with a minor contribution
from photospheric emission of old stars. \citet{Verley2007} argue that
this diffuse emission accounts for $\sim$$2/3$ of all
$24$~\micron\ emission in M33. To isolate $24$~\micron\ emission
originating directly from \hii\ regions, we follow a similar approach
to that used to remove DIG from the \ha\ map. The key difference is
that instead of trying to identify all $24$~\micron\ bright sources by
filtering and applying a cut-off to the $24$~\micron\ emission, we use
the existing locations of \hii\ regions to isolate any local
$24$~\micron\ excess associated with \hii\ regions. We
extinction-correct the DIG-subtracted \ha\ emission using only this
local excess in $24$~\micron\ emission. The total integrated flux at
$24$~\micron\ ($R_{\rm gal} \leq 4.5$~kpc) is $3.92 \times
10^{41}$~\ergpers, the fraction of DIG-subtraced $24$~\micron\ inside
the \hii\ region mask is $1.63 \times 10^{41}$~\ergpers\ ($42$\%). The
the $24$~\micron\ correction implies H$\alpha$ extinctions of $A_{\rm
  H\alpha} \sim 0.3-0.4$ magnitudes.

\section{Methodology}
\label{sec:methodology}

To quantify the scale-dependence of the molecular star formation law,
we measure the \htwo\ depletion time\footnote{We emphasize that
  $\tau_{\rm dep}$ maps directly to observables. It is proportional to
  the ratio of CO to extinction-correct H$\alpha$ emission.},
$\tau_{\rm dep} = \htwosd / \SFRsd$, for apertures centered on bright
CO and \ha\ peaks. We treat the two types of peaks separately and vary
the sizes of the apertures used. In this way we simulate a continuum
of observations ranging from nearly an entire galaxy ( $>1$~kpc
apertures) to studies of (almost) individual GMCs or \hii\ regions
($75$~pc apertures). The CO data limit this analysis to galactocentric
radius $< 4.5$~kpc ($\sim$$0.6$~r$_{25}$).

\subsection{Identifying \co\ and \ha\ Peaks}
\label{subsec:peaks}

We employ a simple algorithm to identify bright regions in the
DIG-subtracted, extinction-corrected H$\alpha$ map and the integrated
CO intensity map. This automated approach allows us to use the same
technique on both maps to find peaks matched in scale to our smallest
aperture ($75$~pc). It is also easily reproducible and extensible to
other galaxies.

This algorithm operates as follows: We identify all contiguous regions
above a certain intensity --- the local $3\sigma$ in the CO map and
$\sim$$1.9 \times 10^{40}$~erg~s$^{-1}$~kpc$^{-2}$ in the corrected
\ha\ map ($\sim$$0.1$~M$_\odot$~yr$^{-1}$~kpc$^{-2}$ following Equation
\ref{eq:sfr}). We reject small regions (area less than $\sim$$110$~arcsec$^2$,
which correspond to $\sim$$1800$~pc$^2$ at the distance of M33)
as potentially spurious; the remaining regions are expanded
by $20$\arcsec\ ($\sim$$80$~pc) in radius to include any low
intensity envelopes. The positions on which we center our apertures
are then the intensity-weighted average position of each distinct
region.

We find $172$ \co\ regions and $154$ \ha\ regions. Strictly speaking,
these are discrete, significant emission features at $\sim$$50$~pc
resolution. At this resolution, there is a close but not perfect match
between these peaks and the real physical structures in the two maps
--- GMCs and \hii\ regions. Figure~\ref{f1} ({\em middle panels})
shows our peaks along with the cataloged positions of GMCs
\citep{Rosolowsky2007} and \hii\ regions \citep{Hodge2002}. There is a
good correspondence, with $>80\%$ of the $149$ known GMCs and the 150
brightest \hii\ regions lying within $\sim$$6$\arcsec\ (3 pixels) of
one of our regions.

\subsection{Measuring Depletion Times}
\label{subsec:depletiontimes}

For a series of scales $d$, we center an aperture of diameter $d$ on
each CO and \ha\ peak and then measure fluxes within that aperture to
obtain a mass of \htwo\ ($M_{\rm H2}$) and a star formation rate
(SFR). We then compute the median \htwo\ depletion time for the whole
set of apertures. We do this for scales $d=1200$, $600$, $300$, $150$ and
$75$~pc and record results separately for apertures centered on CO and
\ha\ peaks.

At larger spatial scales, apertures centered on different peaks overlap
(because the average spacing between CO and H$\alpha$ peaks is
less than the aperture size). To account for this, we measure only
a subset of apertures chosen so that at least $80$\% of the selected
area belongs only to one aperture targeting a given peak type (CO or
\ha) at one time.

While we center on particular peaks, we integrate over all emission in
our maps within the aperture. At the smallest scales we probe
($75$~pc), this emission will arise mostly --- but not exclusively ---
from the target region. At progressively larger scales, we will
integrate over an increasing number of other regions.

\begin{deluxetable}{cccc}
\tablecolumns{4}
\tablecaption{$\tau_{\rm dep}$ as Function of Peak and Scale \label{tab:tdep}}
\tablehead{\multicolumn{1}{c}{Scale} &
	\multicolumn{2}{c}{Depletion Time (Gyr)} \\
	\multicolumn{1}{c}{(pc)} &
	\multicolumn{1}{c}{centered on CO} &
	\multicolumn{1}{c}{centered on H$\alpha$} & $\left<N\right>$\tablenotemark{a}}
\startdata
	1200		& $1.1 \pm 0.1$ & $0.9 \pm 0.1$ & 16.2 \\
	600		& $1.2 \pm 0.3$ & $1.0 \pm 0.1$ & 5.2 \\
	300		& $2.5 \pm 0.5$ & $0.64 \pm 0.05$ & 2.1 \\
	150		& $4.9 \pm 0.9$ & $0.41 \pm 0.04$ & 1.4 \\
	75		& $8.6 \pm 2.1$ & $0.25 \pm 0.02$ & 1.1
\enddata
\tablenotetext{a}{Typical number of individual CO or H$\alpha$ peaks inside an aperture.}
\end{deluxetable}

\subsection{Uncertainties}
\label{subsec:montecarlo}

We estimate the uncertainty in our measurements using a Monte-Carlo
analysis. For the high-SNR \ha\ and $24$~\micron\ maps, we add
realistic noise maps to the observed ``true'' maps and repeat the
identification and removal of DIG emission using smoothing kernels and
emission measure cuts perturbed from the values in
Section~\ref{subsec:hadata} by $\pm 25$\%. The low SNR of the CO data
requires a more complex analysis.  We assume that all regions with
surface densities above $10$~\Msunperpc\ ($\sim$$1.5$ $\sigma$) in the
integrated CO map contain true signal. We generate a noise map
correlated on the ($13\arcsec$) spatial scale of our CO data and scale
this noise map according to the square root of the number of channels
along each line of sight in our masked CO cube (typically $5-7$). Then
we add all emission from the pixels above $10$~\Msunperpc. Finally, we
re-identify peaks in the new maps and re-measure $M_{\rm H2}$ and SFR
in each region. We repeat this process $100$ times; the scatter in
$\tau_{\rm dep}$ across these repetitions is our uncertainty estimate.

\section{Results}
\label{sec:results}

\begin{figure}
	\epsscale{1.0}
	\plotone{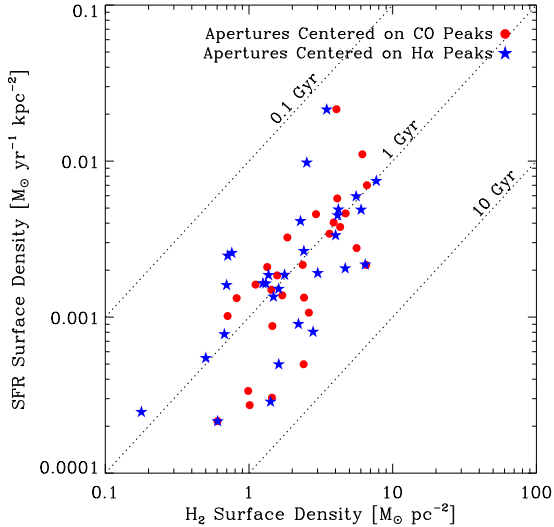}
	\caption{The relationship between SFR surface density ($y$-axis)
	and H$_2$ surface density ($x$-axis) at large spatial scales. Red
	points show measurements integrated over (one independent
	subset of) 1200 pc apertures centered on CO peaks. Blue points
	show similar measurements centered on H$\alpha$ peaks. Dashed
	lines indicate fixed H$_2$ depletion times ranging from $0.1$~Gyr
          in the upper left to $10$~Gyr in the lower right.\label{f2}}
\end{figure}

Figure \ref{f2} shows a well-known result for M33. There is a strong
correlation (rank correlation coefficient of $r \approx 0.8$) between
the surface densities of SFR and \htwo\ at 1200 pc scales. Power law
fits to the different samples (types of peaks) and Monte Carlo iterations yield
\htwo\ depletion times, $\tau_{\rm dep} = M_{\rm H2} / {\rm SFR}$, of
$\sim$$1$~Gyr (at \htwo\ surface densities of $3$~\Msunperpc) and power
law indices of $\sim$$1.1 - 1.5$.  These results (modulo some
renormalization due to different assumptions) match those of
\citet{Heyer2004} and \citet{Verley2010} in their studies of the
star formation law in M33. The important point here is that there is
good evidence for an internal H$_2$-SFR surface density relation in M33.

\begin{figure}
	\epsscale{1.0}
	\plotone{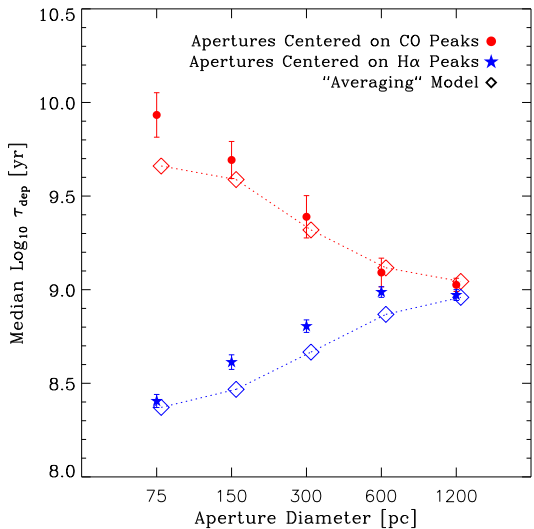}
	\caption{Scale dependence of the \htwo\ depletion time,
          $\tau_{\rm dep}$, in M33. The $y$-axis shows the logarithm
          of the median \htwo\ depletion time for apertures of
          different diameters ($x$-axis) centered on CO peaks (red)
          and H$\alpha$ peaks (blue). Error bars correspond to
          uncertainty in the median estimated via a Monte-Carlo
          analysis. Dashed lines show expectation for simply averaging
          together two populations of regions in different states
          (Section \ref{sec:disc}).\label{f3}}
\end{figure}

We plot the median $\tau_{\rm dep}$ as a function of scale (aperture
size) in Figure~\ref{f3}, giving results for apertures centered on CO
(red circles) and H$\alpha$ (blue stars) peaks seperately. For the
largest scales, we find a similar $\tau_{\rm dep}$ for both sets of
apertures (as was evident from Figure \ref{f2}). Going to smaller
aperture sizes, $\tau_{\rm dep}$ becomes a strong function of scale
and type of peak targeted. Small apertures centered on CO peaks have
very long $\tau_{\rm dep}$ (up to 10 Gyr). Small apertures targeted
toward H$\alpha$ peaks have very short $\tau_{\rm dep}$
($0.3$~Gyr). This may not be surprising, given the expectations that
we outlined in Section~\ref{sec:intro} and the distinctness of the
bright \ha\ and CO distributions seen in the lower left panel of
Figure~\ref{f1}, but the dramatic difference as one goes from
$\sim$kpc to $\sim$$100$~pc scales is nonetheless striking.

A few caveats apply to Figure~\ref{f3}. First, in subtracting the
diffuse emission (DIG) from the \ha\ map, we removed $\sim$$40\%$
of the flux. This could easily include faint regions associated with CO
peaks, which instead show up as zeros in our map.  Perhaps more
importantly, we use the $24$~\micron\ map {\em only} to correct the
DIG-subtraced \ha\ map for extinction. Any completely embedded star
formation will therefore be missed. For both of these reasons, the SFR
associated with the red points, while it represents our best guess,
may be biased somewhat low and certainly reflects emission from
relatively evolved regions --- those regions that have \ha\ fluxes
above our DIG-cutoff value. There is no similar effect for the CO map.

\begin{figure}
	\epsscale{1.0}
	\plotone{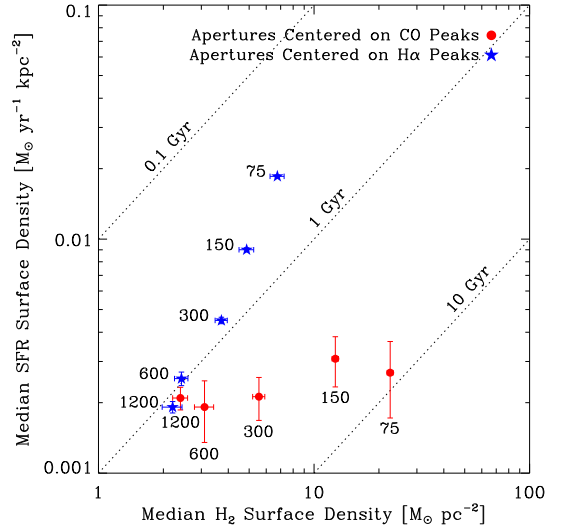}
	\caption{Scale dependence of the location of data in the
          star formation law parameter space. Red points show the
          median $\Sigma_{\rm SFR}$ ($y$-axis) and $\Sigma_{\rm H2}$
          ($x$-axis) for apertures centered on CO peaks. Blue stars
          show the same for apertures centered on H$\alpha$ peaks.
          Dashed lines as in Figure~\ref{f2}.\label{f4}}
\end{figure}

Figure \ref{f3} implies that there is substantial movement of points
in the star formation law parameter space as we zoom in to higher
resolution on one set of peaks or another. Figure \ref{f4} shows this
behavior, plotting the median $\Sigma_{\rm SFR}$ and median
$\Sigma_{\rm H2}$ for each set of apertures (N.B., the ratio of median
$\Sigma_{\rm H2}$ to median $\Sigma_{\rm SFR}$ does not have to be
identical to the median $\tau_{\rm dep}$; the difference is usually
$\lesssim 30\%$). We plot only medians because individual data are
extremely uncertain, include many upper limits, and because we are
primarily interested in the systematic effects of resolution on data
in this parameter space.

Apertures centered on CO peaks (red points) have approximately
constant \SFRsd , regardless of resolution. This can be explained if
emission in the \ha\ map is homogeneously distributed as compared to the
position of CO peaks. Meanwhile there is a strong change in \htwosd\ for
decreasing aperture sizes on the same peaks; $\Sigma_{\rm H2}$ goes up
as the bright region fills more and more of the aperture. A similar
effect can be seen for the H$\alpha$ (blue stars), though there
is more evolution in \htwosd\ with increasing resolution because most
bright H$\alpha$ regions also show some excess in CO emission.

\section{Discussion}
\label{sec:disc}

Figure \ref{f3} shows that by zooming in on an individual star-forming
region, one loses the ability to recover the star formation law
observed on large scales. For apertures $\lesssim 300$~pc in size, the
relative amounts of CO emission and \ha\ intensity vary systematically
as a function of scale and what type of region one focuses on. Another
simple way to put this, demonstrated in Figure \ref{f4}, is that
scatter orthogonal to the SFR-H$_2$ relation increases with increasing
resolution. Eventually this washes out the scaling seen on large
scales and the star formation law may be said to ``break down''.

What is the origin of this scale dependence? In principle one can
imagine at least six sources of scale dependence in the star formation
law:

\begin{enumerate}
\item Statistical fluctuations due to noise in the maps.
\item Feedback effects of stars on their parent clouds.
\item Drift of young stars from their parent clouds.
\item Region-to-region variations in the efficiency of star formation.
\item Time-evolution of individual regions.
\item Region-to-region variations in how observables map to physical
  quantities.
\end{enumerate}

Our observations are unlikely to be driven by any of the first three
effects. In principle, statistical fluctuations could drive the
identification of H$\alpha$ and CO peaks leading to a signal similar
to Figure \ref{f3} purely from noise. However, our Monte Carlo
calculations, the overall SNR in the maps, and the match to previous
region identifications make it clear that this is not the case.

Photoionization by young stars can produce CO shells around
\hii\ regions inside of a larger cloud or complex. This is a clear
case of a small-scale offset between H$\alpha$ and CO. However, the
physical scales in Figure \ref{f3} are too large for this effect to
have much impact, it should occur on scales more like $\sim$$10$~pc.

Similarly, the scales over which $\tau_{\rm dep}$ diverges between CO
and H$\alpha$ peaks ($75-300$ pc) are probably too large to be
produced by drift between young stars and their parent cloud. A
typical internal GMC velocity dispersion in M33 is a few km~s$^{-1}$
\citep[$1\sigma$;][]{Rosolowsky2003}. Over an average cloud lifetime
\citep[$\sim$$30$~Myr;][]{Blitz2007,Kawamura2009}, this implies a drift
of at most $100$~pc. This extreme case is just large enough to register
in our plot but unlikely to drive the signal we see at scales of
$150-300$ pc. See \citet{Engargiola2003} for a similar consideration
of GMCs and \hi\ filaments.

Instead of drifts or offsets, what we observe is simply a lack of
direct correspondence between the CO and \ha\ luminosities of
individual star-forming regions. The brightest $150$ CO peaks are
simply not identical to the brightest $150$ \ha\ peaks. The bottom
right panel of Figure \ref{f1} shows this clearly; about a third of the
peaks in M33 are nearer to another peak of their own type (i.e., CO to
CO or H$\alpha$ to H$\alpha$) than to a peak of the other type. Thus
Figure \ref{f3} shows that the ratio of CO to \ha\ emission varies
dramatically among star-forming regions. In this case the size scale
on the $x$-axis in Figure~\ref{f3} is actually a proxy for the number
of regions inside the aperture. In M33, apertures of $75$~pc diameter
usually contain a single peak. At $150$~pc, this is still the case
$\sim$$70\%$ of the time, and at $300$~pc only a few regions
are included in each aperture.

Why does the ratio of CO-to-H$\alpha$ vary so strongly from
region-to-region? The efficiency with which gas form stars may vary
systematically from region to region (with high H$\alpha$ peaks being
high-efficiency regions), star-forming regions may undergo dramatic
changes in their properties as they evolve (with H$\alpha$ peaks being
evolved regions), or the mapping of observables to physical quantities
(Equations \ref{eq:xco} and \ref{eq:sfr}) may vary from region to
region.

It is difficult to rule out region-to-region efficiency variations,
but there is also no strong evidence for them. \citet{Leroy2008}
looked for systematic variations in $\tau_{\rm dep}$ as a function of
a number of environmental factors and found little evidence for any
systematic trends. \citet{Krumholz2005, Krumholz2007} suggested that
the cloud free-fall time determines $\tau_{\rm dep}$ to first order,
but based on \citet{Rosolowsky2003}, the dynamic range in free-fall
times for M33 clouds is low. On the other hand, \citet{Gardan2007}
found unusually low values of $\tau_{\rm dep}$ in the outer disk of
M33.

There is strong evidence for evolution of star-forming regions.
\citet{Fukui1999}, \citet{Blitz2007}, \citet{Kawamura2009}, and
\citet{Chen2010} showed that in the LMC, the amount of H$\alpha$
and young stars associated with a GMC evolves significantly across
its lifetime. In our opinion this is the most likely explanation for the
behavior in Figure \ref{f3}. Star-forming regions undergo a very
strong evolution from quiescent cloud, to cloud being destroyed by
\hii\ region, to exposed cluster or association.  When an aperture
contains only a few regions, $\tau_{\rm dep}$ for that aperture will
be set by the evolutionary state of the regions inside it. That state
will in turn determine whether the aperture is identified as a CO peak
or an H$\alpha$ peak. CO peaks will preferentially select sites of
heavily embedded or future star formation while \ha\ peaks are
relatively old regions that formed massive stars a few Myr ago.

Region-to-region variations in the mapping of observables (CO and
H$\alpha$) to physical quantities (H$_2$ mass and SFR) are expected.
Let us assume for the moment that the ratio of H$_2$ to SFR is
constant and independent of scale. Then to explain the strong scale
dependence of the ratio of CO to H$\alpha$ in Figure \ref{f3}
there would need to be much more \htwo\ per unit CO near H$\alpha$
peaks and many more recently formed stars per ionizing photon near
the CO peaks. At least some of these effects have been claimed: e.g.,
\citet{Israel1997} find a strong dependence of \xco\ on radiation field
and \citet{Verley2010} suggest that incomplete sampling of the IMF
in regions with low SFRs drive the differences they observe between
star formation tracers. However, both claims are controversial and
it seems very contrived to invoke a scenario where only this effect
drives the breakdown in Figure \ref{f3}. It seems more plausible that
the mapping of observables to physical quantities represents a
secondary source of scatter correlated with the evolutionary state
of a region (e.g., the age of the stellar population).

\subsection{Comparison to a Simple Model}
\label{subsec:model}

We argue that the behavior seen in Figures \ref{f3} and \ref{f4} comes
from averaging together regions in different states. Here we implement
a simple model to demonstrate that such an effect can reproduce the
observed behavior.

The model is as follows: we consider a population of regions. We
randomly assign each region to be an ``\ha\ peak'' or a ``CO peak''
with equal chance of each. CO peaks have $5$ times as much CO as
\ha\ and \ha\ peaks have $5$ times as much \ha\ as CO (roughly driven
by the difference between the results for $75$~pc apertures in
Table~\ref{tab:tdep}). Physically, the idea is simply to build a
population of regions that is an equal mix ``young'' (high CO-to-\ha)
and ``old'' (low CO-to-\ha). Dropping an aperture to contain only a
young (CO peak) or old (H$\alpha$ peak) region will recover our
results at $75$~pc scales by construction. Next we average each of our
original region with another, new region (again randomly determined to
be either a CO or \ha\ peak). We add the CO and \ha\ emission of the
two region together, record the results. We then add a third region
(again randomly young or old), and so on.

The result is a prediction for the ratio of CO to \ha\ as a function
of two quantities: 1) the number of regions added together and 2) the
type of the first region (CO or \ha\ peak). Using the average number of
regions per aperture listed in Table~\ref{tab:tdep} and normalizing to
an average depletion time of 1 Gyr, we then have a prediction for
$\tau_{\rm dep}$ as a function of scale. This appears as the diamond
symbols and dashed lines in Figure~\ref{f3}.

Given the simplicity of the model, the agreement between observations
and model in Figure \ref{f3} is good. Our observations can apparently
be explained largely as the result of averaging together star-forming
regions in distinct evolutionary states. At scales where a single
region dominates, the observed $\tau_{\rm dep}$ is a function of the
state of that region. As more regions are included, $\tau_{\rm dep}$
just approaches the median value for the system.

\subsection{$\tau_{\rm dep}$ vs. Scale at Different Radii}
\label{subsec:rad}

The star formation law apparently breaks down (or at least includes a
large amount of scatter) on scales where one resolution element
corresponds to an individual star-forming region. The spatial
resolution at which this occurs will vary from system to system
according to the space density of star-forming regions in the system.

The surface densities of star formation and \htwo\ vary with radius in
M33 \citep[][]{Heyer2004}. This allows us to break the galaxy into two
regions, a high surface density inner part ($r_{\rm gal} = 0 - 2.2$~kpc)
and a low surface density outer part ( $r_{\rm gal} = 2.2 - 4.5$~kpc).
We measure the scale dependence of $\tau_{\rm dep}$ for each region
in the same way that we did for all data. An important caveat is that
the DIG subtraction becomes problematic for the outer region,
removing a number of apparently real but low-brightness \hii\ regions
from the map. We achieve the best results for large radii with the
DIG subtraction is turned off and report those numbers here. The basic
result of a larger-scale of divergence in the outer disk remains
the same with the DIG subtraction on or off.

We find the expected result, that $\tau_{\rm dep}$ for CO and
H$\alpha$ peaks diverges at larger spatial scales in the outer disk
than the inner disk. In both cases the ratio of $\tau_{\rm dep}$ at CO
peaks to $\tau_{\rm dep}$ at H$\alpha$ peaks is $\sim$$1$ for
$1200$~pc apertures. For $600$~pc apertures that ratio remains
$\sim$$1$ in the inner disk but climbs to $\sim$$2$ in the outer disk,
suggesting that by this time there is already some breakdown in the
SFR-H$_2$ relation. For $300$~pc apertures, the same ratio is
$\sim$$1.7$ in the inner disk and $\sim$$3$ in the outer disk. It thus
appears that at large radii in M33 the star formation law breaks down
on scales about twice that of the inner disk, though the need to treat
the DIG inhomogeneously means that this comparison should not be
overinterpreted.

\section{Conclusions}
\label{sec:conc}

Our main conclusion is that the molecular star formation law observed
in M33 at large scales \citep[e.g.,][]{Heyer2004,Verley2010} shows
substantial scale dependence if one focuses on either CO or
\ha\ peaks. The median depletion time (or CO-to-H$\alpha$ ratio)
measured in a $75$~pc diameter aperture (derived from averaging
$\sim$$150$ such apertures) varies by more than an order of magnitude
between CO and \ha\ peaks. At large ($\sim$kpc) scales this difference
mostly vanishes. We argue that the scale for the breakdown is set by
the spatial separation of high-mass star-forming regions, with the
breakdown occurring when an aperture includes only a few such regions
in specific evolutionary states (a scale that corresponds to
$\sim$$300$~pc in M33).

In this case the scaling relation between gas and star formation rate
surface density observed at large scales does not have its direct
origin in an instantaneous cloud-scale relation. Individual GMCs and
\hii\ regions will exhibit a CO-to-H$\alpha$ ratio that depends on
their evolutionary state (likely with significant additional
stochasticity) and as a result the $\sim$$150$ brightest objects at a
given wavelength will be a function of what evolutionary state that
observation probes. This divergence is consistent with recent results
from the LMC \citep{Kawamura2009} indicating that individual GMCs
exhibit a range of evolutionary states over their $20 - 30$ Myr
lifetime.

This does not mean that comparisons of tracers of recent and future
star formation on small scales are useless. To the contrary, such
observations contain critical information about the evolution of
individual clouds as a function of time and location that is washed
out at large scales ($\gtrsim 300$~pc in M33). However, once one moves
into the regime where a single object contributes heavily to each
measurement, it is critical to interpret the results in light of the
evolution of individual clouds.

\acknowledgements We thank Mark Heyer and Edvige Corbelli for
providing us the \co\ data, Rene Walterbos for the \ha\ image, and
Karl Gordon for the \emph{Spitzer} images of M33. We thank the
anonymous referee for a careful and constructive report that improved
the paper. We acknowledge the use of the \hii\ region catalog from
Hodge et al. AKL thanks Michele Thornley and Jack Gallimore for
helpful discussions.  The work of \mbox{A. S}. was supported by the
Deutsche Forschungsmeinschaft (DFG) Priority Program 1177. This
research made use of the NASA/IPAC Extragalactic Database (NED), which
is operated by the JPL/Caltech, under contract with NASA as well as
NASA's Astrophysical Data System (ADS). Support for AKL was provided
by NASA through Hubble Fellowship grant HST-HF-51258.01-A awarded by
the Space Telescope Science Institute, which is operated by the
Association of Universities for Research in Astronomy, Inc., for NASA,
under contract NAS 5-26555.

\begin{appendix}

Here we test how our method of selecting peaks and the removal of a
diffuse ionized component affect our results.

\begin{figure*}
	\epsscale{1.0}
	\plottwo{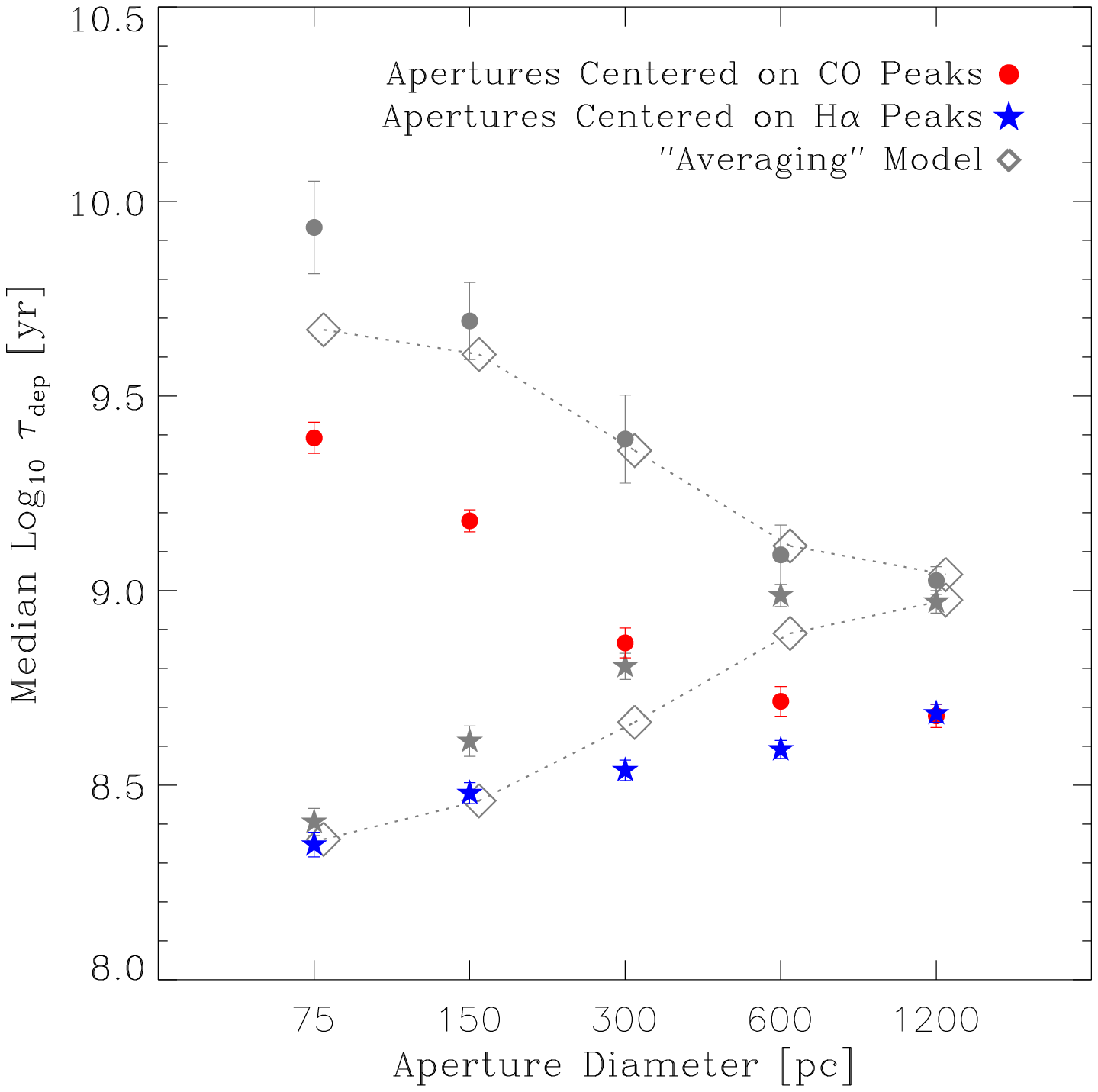}{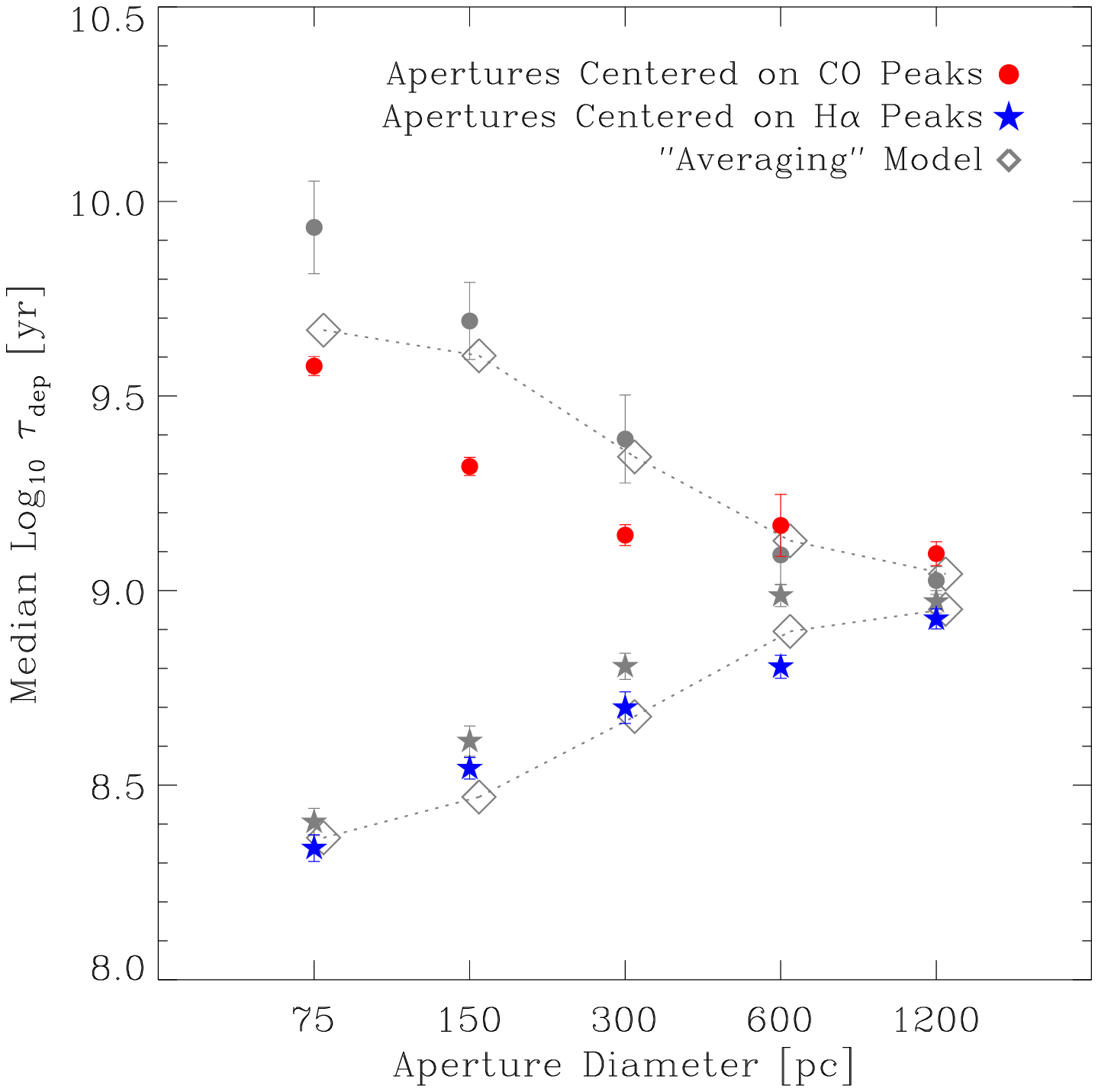}
	\caption{\label{f5} Tests of systematics. We
          repeat the measurement of $\tau_{\rm dep}$ and remake Figure
          \ref{f3} varying our methodology. {\em Left:}
          Results if we perform no DIG-subtraction on the H$\alpha$
          maps. {\em Right:} Results if we use the known positions of
          GMCs and \hii\ regions from \citet{Rosolowsky2007} and
          \citet{Hodge2002}. In both panels, our original results
          using the DIG-subtracted maps and an automated peak finder
          appear as gray points. Our qualitative conclusions are
          largely robust to these changes in methodology, though the
          overall normalization of CO-to-H$\alpha$ ratio does depend
          on the DIG subtraction.}
\end{figure*}

First, we repeating the analysis on the original maps without any
DIG subtraction. We show the results in the left panel of Figure~\ref{f5},
along with the original measurements (in gray) from Figure~\ref{f3}.
To first order, the scale-dependence of $\tau_{\rm dep}$ is unchanged,
but the \htwo\ depletion are offset; $\tau_{\rm dep}$ derived from maps
with no DIG subtraction is a factor of $\sim$$2$ smaller from the 
DIG-subtracted maps.

As a second test we assess the impact of the particular choice of
aperture positions. In main text we used a peak-finding
algorithm. Here we test the effect of using published positions of
GMCs and \hii\ regions instead. The right panel in Figure~\ref{f5}
shows our original data in gray while the red and blue points are
$\tau_{\rm dep}$ derived using the \citet{Rosolowsky2007} and
\citet{Hodge2002} catalogs. The median $\tau_{\rm dep}$ at large scale
is unchanged from Figure~\ref{f3} (gray values). However, for
apertures centered on GMCs, $\tau_{\rm dep}$ on small scales does
change from our analysis. This difference originates in different
numbers and locations of the positions that are studied.  In our
original analysis, we study $172$ positions which have CO emission
peaks above $3\sigma$. The Rosolowsky catalog, on the other hand,
consists of only $140$ positions inside a galactocentric radius of
$4.5$~kpc. In addition, a subset of the two samples targets different
regions in M33: First, the catalog positions tend to be more clustered
than the ``peak'' positions which leads to a somewhat larger number
($5 - 15\%$) of objects in the smaller apertures and a smaller deviation
in depletion times for CO or \hii\ centered apertures.  Second, while
the molecular gas surface densities at the positions of the two
samples do not differ significantly, the star formation rate surface
densities are a factor of $\sim$$3$ higher for the catalog positions as
compared to the (more numerous) ``peak'' positions.  This leads to
shorter \htwo\ depletion times on small scales for the catalog sample.

Both tests show that the analyzed scale dependence of the star
formation relation and the determination of its origin is not strongly
dependent on the particular methodology chosen in this paper. While
global shifts in the derived depletion times can arise due to the
subtraction of diffuse \ha\ emission, we find only small variations in
the scale dependence due to different selection of positions where we
perform our study.

\end{appendix}

\end{document}